# Reliable multicast using fault tolerant MPI in the Grid environment

Benoit Hudzia, University College of Dublin
And Serge G. Petiton, Laboratoire d'Informatique Fondamentale de Lille
Emails: Benoit.hudzia@ucd.ie, Petiton@lifl.fr

*Abstract:* **Grid environments have recently been developed with low stretch and overheads that increase with the logarithm of the number of nodes in the system. Getting and sending data to/from a large numbers of nodes is gaining importance due to an increasing number of independent data providers and the heterogeneity of the network/Grid. One of the key challenges is to achieve a balance between low bandwidth consumption and good reliability. In this paper we present an implementation of a reliable multicast protocol over a fault tolerant MPI: MPICH-V2. It can provide one way to solve the problem of transferring large chunks of data between applications running on a grid with limited network links. We first show that we can achieve similar performance as the MPICH-P4 implementation by using multicast with data compression in a cluster. Next, we provide a theoretical cluster organization and GRID network architecture to harness the performance provided by using multicast. Finally, we present the conclusion and future work.**

## I. INTRODUCTION

When 1 to N communication takes place in Grid Environments, it places a heavy burden on network links. To provide a partial solution on this issue we will present an implementation of a reliable multicast over a fault tolerant MPICH-V2 [1] which is part of the Xtremweb project [2]. We will first show that we can achieve similar performance of MPICH-P4 [3] implementation by using the multicast with data compression within a Grid created by using XtremWeb. Next we provide two theoretical Grid network architectures to harness the performance provided by using multicast. Finally, we present the conclusion and future work.

## II. MULTICAST IMPLEMENTATION OF MPICH-V2 / XTREMWEB

To provide multicast into MPI applications running in Grid we have decided to implement those functionalities in MPICH-V2 and Xtrem-Web because the conjunction of both provide most of the functionalities of grid environments. To avoid redeveloping an efficient and reliable multicast protocol from scratch we decided to use MCL library [4] that can provide multicast with ACL [5], or NORM [6] protocol and with FEC [7] decoding.

### A. MPICH-V2 and EXTREMWEB

MPICH-V2 implements a pessimistic sender-based protocol on top of MPICH 1.2.5 [16] (with the help of Condor [8] for fault tolerance), using a dispatcher, a checkpoint scheduler, some event loggers, checkpoints servers, computing nodes and their communication daemons. The figure 1 presents a typical setup of a running MPICH-V2 system, where the dispatcher, the event logger and the checkpoint scheduler running on the same computer.

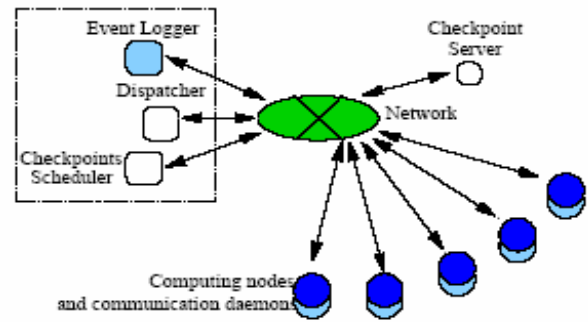

Figure 1. MPICH-V2 architecture

MPICH-V2 can be implemented within XtremWeb (XW). XtremWeb's main goal, as a Global Computing platform, is to compute distributed applications using idle time of widely interconnected machines. In conjunction with MPICH-V2, XtremWeb provides scalability and fault tolerance to applications and users. It also allows the adding and removing of nodes on the fly and permits them to exist in separate locations. These features can be summed up in one phrase a "Computing Grid", as show in figure 2.

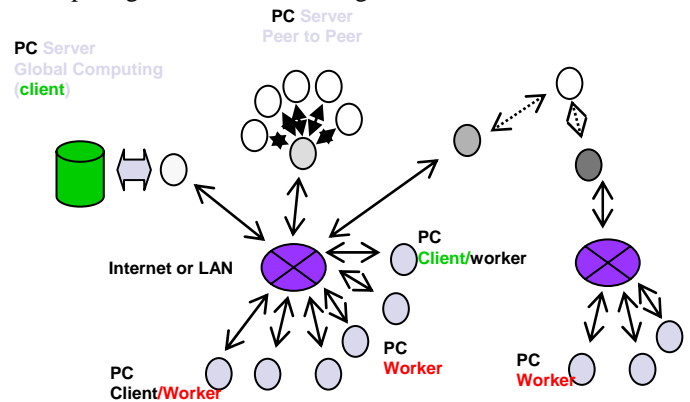

Figure 2. XtremWeb architecture

## B. Multicast library (MCL):

### 1) Asynchronous layered coding (ALC/LCT) Protocol

The MCL library gives us two kinds of multicast protocol, ALC and NORM. ALC is based on the concept of layered coding that was first introduced with reference to audio and video streams. LCT provides transport level support for massively scalable protocols using the IP multicast network service. An LCT session comprises multiple channels originating at a single sender that are used for some period of time to carry packets pertaining to the transmission of one or more objects that can be of interest to receivers as shown in figure 3. Two of the key difficulties in scaling reliable content delivery using IP multicast are dealing with the amount of data that flows from receivers back to the sender, and the associated response (generally data retransmissions) from the sender. Protocols that avoid any such feedback, and minimize the amount of retransmissions, can be massively scalable. LCT can be used in conjunction with FEC codes or a layered codec to achieve reliability with little or no feedback. This can be very useful in a large cluster network or in Grid system with different connection linking each computing centre.

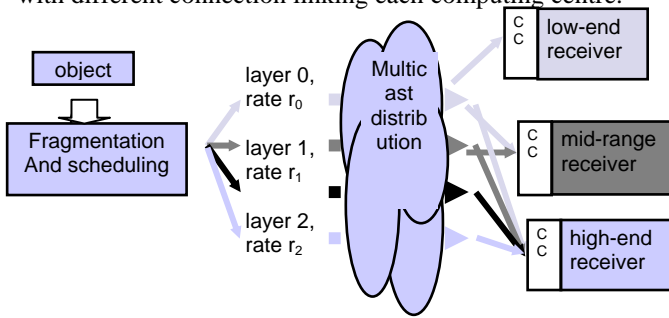

Figure 3. ALC/LCT

### 2) NACK Oriented Reliable Multicast (NORM) protocol

NORM is the second protocol integrated in MCL, the Negative-acknowledgement (NACK) Oriented Reliable Multicast (NORM) protocol is designed to provide reliable transport of data from one or more sender(s) to a group of receivers over an IP multicast network. The primary design goals of NORM are to provide efficient, scalable, and robust bulk data (e.g., computer files, transmission of persistent data) transfer across possibly heterogeneous IP networks and topologies. However, due to the instability of the NORM protocol in the library, this protocol was not tested.

### 3) FEC: Forward Error Coding

FEC codes are a valuable basic component of any transport protocol that is to provide reliable delivery of content [9]. The great advantage of using FEC with multicast transmissions is that the same parity packet can recover different lost packets at different receivers. Furthermore, FEC codes can ameliorate or even eliminate the need for feedback from receivers to senders to request retransmission of lost packets as shown in figure 4. Therefore, it can help avoiding a feedback implosion [10].

But it comes with a price. With FEC we have to send 100% to 200% more data. Moreover, building the parity bloc with Reed-Solomon erasure code (RSE) has a huge encoding/decoding time [11] due to a heavy CPU consumption. In addition, this has a direct impact on performance. The creators of MCL are working on a new way of building parity bloc for large data block, but it's still a work in progress.

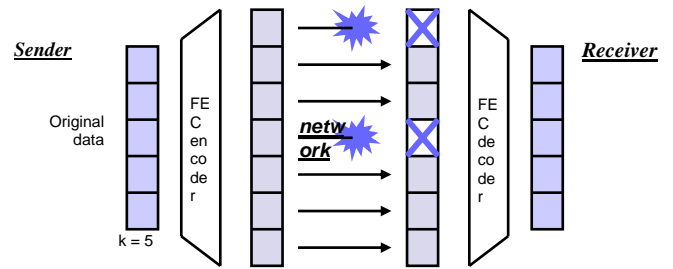

Figure 4. FEC coding / decoding

## C. Multicast

Due to the use of condor libraries for the fault tolerance the use of threads in MPICH-V2 is forbidden. We decided to outsource the MCL threads in the communication daemon of V2. This one is independent and is not check-pointed. We just spawn a thread at the start-up of the communication daemons, which is going to contain the MCL threads for multicast communication. All the communications between the multicast process and MPICH-V2 will exist through a shared memory segment. To avoid memory consumption and spawning a new memory segment for every transfer of data we just have created a single segment of memory, if the data buffer is too large we just broke it into pieces that fit our buffer for transmission. The data marshalling, and compression still take place in the main MPICH-v2 process.

We encountered other difficulties during the implementation: Buffer in the MCL library. When we pass data to the MCL library and modify it before it has built the entire block that have to be sent, there is a chance that MCL will work on the modified data. That is because the mcl_send function is non-blocking. If mcl_send was blocking the program would have to wait for a very long time for the completion of the transfer even if the receiver had already received all the data. These effects are due to the combination of the connectionless protocol (UDP) and the multi-channel protocol (ALC) which can transfer data at a very low rate using a low data rate transfer channel. Due to the use of shared memory to transfer data between MPI and the communication daemon we have to make sure that the communication library does not process corrupted data, i.e. the common shared memory problem.

To avoid recreating a new multicast stream when a node replays all the communication due to a restart after a failure, the multicast communications are treated as unicast communications. The data will be requested from the originating node and directly transferred from it. Further improvement on this part can be made, for example by requesting the data from any node existing in the group that received it.

III. EVALUATION OF MULTICAST

A. *Simple multicast and comparison with P4 and V2 performance*

The experimental tests presented in the following section have been run on a low-end cluster of PCs under Linux 2.4.18 with slow network connections. The cluster has been used in dedicated mode to ensure a fair comparison between the different implementations. The cluster consists of two major parts: 32 computing nodes and 12 auxiliary machines connected to a single 48 port Ethernet 100 Mbits switch. The computing nodes are equipped with Athlon XP 1800+, running at 1.5 GHz and 1GByte of main memory plus 1GBytes of swap on IDE disk. The second part of the cluster is composed of dual-Pentium III machines, with processors running at 500MHz with 512MByte of main memory and 1GBytes of swap on IDE disk and a third part composed of Athlon XP 2200+, running at 1.73 GHz and 1 GByte of main memory plus 1GByte of swap on IDE disk. All MPI implementations use MPICH 1.2.5 version. Test programs were compiled using the GNU GCC.
For this experiment, we used 400 different matrixes from the matrixarket [17] web site because matrix computation in cluster relies heavily on sending large chunk of data to multiple nodes. Note: a throughput represent the time to transfer data including the encoding/decoding time (compression, FEC, etc…).

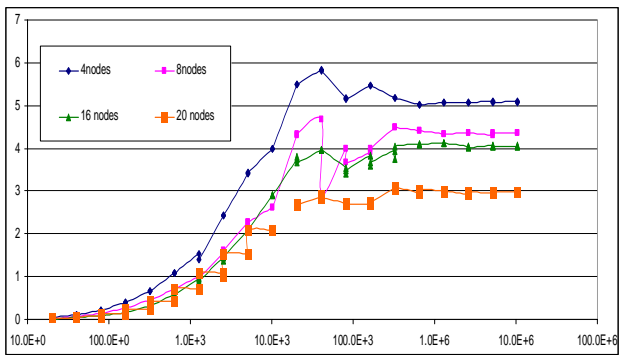

Figure 5. Mpich-P4 throughput (Mb/s) Vs matrix size (Byte)

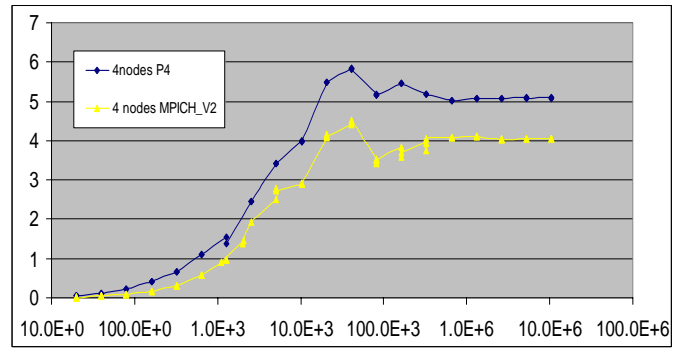

Figure 6. MpichV2 an P4 Data throughput (Mb/s) Vs matrix size (byte)

The MPI_bcast is done in MPICH by using a binomial tree. This means for every power of two nodes the efficiency decreases. This algorithm is one of the best since the case of a unicast implemented broadcast with a known topology, but it is obliged to send the data N times which consumes a lot of bandwidth, and bandwidth is one of the major resources we lack. Data transportation takes longer than computation. The graph in figure 5 shows the impact on the throughput when the number of nodes vary. We clearly see that when the threshold (a power of two) for the number of nodes is reached the throughput drops.

MPICH-V2 is slightly slower than MPICH-P4, as shown in figure 6, but remains always close to MPICH-P4. The difference is explained by the acknowledgement of message logging with the event loggers [1].

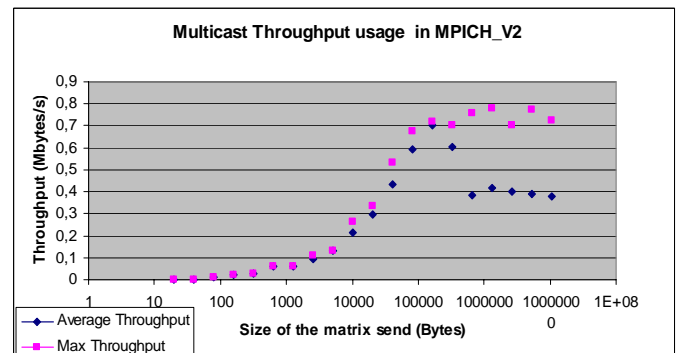

Figure 7 Multicast and Mpich-V2 throughput

The multicast in MPICH-V2 gives very poor results, as we can see in figure 7. We obtain a peak throughput of 0.8 Mega bytes per second. The low performance is mostly due to the overhead of FEC computation, it adds, in this configuration a 200% increase of the data transferred and is time consuming. On top of that, the communication speed is quite low due to the multiplexing of the ALC channels and also because of the artificial congestion protocol implemented (to avoid flooding the network by sending the packets to fast, it is still based on UDP).

In figure 7, we notice that when we start sending matrices of size 140Kb the performance decreases to reach an almost constant level of throughput of 0,38 Mbytes/s. This drop of 50% of performance is due to the size of the reception and transmission buffer linked to the socket. When you send data over a certain size the reception socket fills faster due to the amount of data and you have to wait to get it empty before refilling it. In MCL the size of the buffer is 64Kb. That's also why we reach maximum throughput for 64Kb of data. The maximum throughput that we reach for higher size of the matrix is due to the variation of the degree of complexity for calculation of the FEC data, this one depends on the data contain by the matrix; sometimes the data of these matrixes allow a fast computation of the FEC. Nevertheless, the impact is still great: The throughput is ~ 50 % less than the peak.

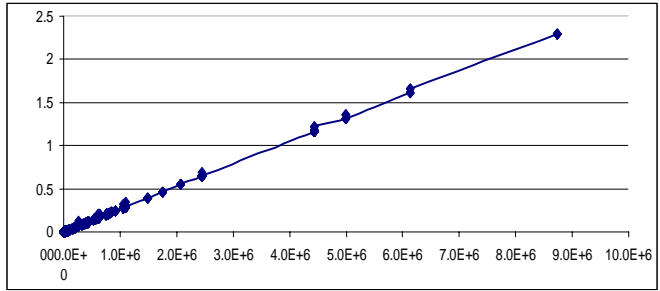

Figure 8. MpichV2 time to transfer data (s) Vs Matrix size (Byte)

Figure 8 shows that the ratio time / size of the data transferred with the multicast version of MPICH-V2, it has a linear growth. That means that data size matters more than the data itself.

But one thing we have to see is that the bandwidth consumption is quite low, with P4 or V2 we have to send N times the matrix, here with the multicast implementation we just need to send the matrix once, with the FEC data. On the other hand, the use of ALC multiplies the bandwidth consumption due to the multi-channel architecture. We have in conclusion for P4 and V2: "N*m" and for V2 multicast: "m+2m * (Number of channels)" which is few when the number of nodes is high (N = number of nodes, m = size of the message).

Also one thing that we have expected and the experienced confirmed it, performance are not modified by the numbers of nodes but more by the performance of the machines used. The heavy CPU use for the calculation of FEC dominates performance. The data to create the graphs in figures 7 and 8 was obtained by measuring the time required to send a matrix a hundred times and with various configurations (from 2 to 24 nodes). The results were similar and independent of the configuration and number of nodes.

### B. Adding compression

#### 1) Data compression

In the implementation of multicast in MPICH-V2, we used the Zlib library [12] to reduce the effect of the FEC computation and also reduce the data to send. This library allows us to use a lossless compressed data format that compresses data using a combination of the LZ77 algorithm and Huffman coding, with efficiency comparable to the best currently available general-purpose compression methods. The data can be produced or consumed, even for an arbitrarily long sequentially presented input data stream, using only an *apriori-bounded* amount of intermediate storage. On top of that, zlib is independent of CPU type, operating system, file system, and character set, and hence can be used for inter-exchange.

Empirically, zlib algorithm is capable of compression factors exceeding 1000:1. (The test case was a 50MB file filled with zeros; it compressed to roughly 49 KB.) More typical zlib compression ratios are approximately 2:1 to 5:1. As shown in the following graph ( figure 9) , the data we have recorded during the execution of the test of the multicast mpich-V2 with data compression and the compression algorithm set to best compression (level 9) on matrices from the matrix market web site[17].

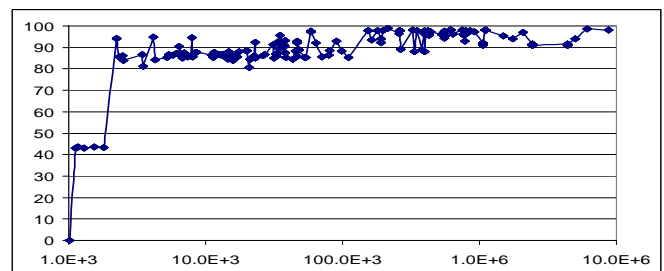

Figure 9. Compression ratio (%) Vs Matrice size

As we can see in figure 9, with the matrix from matrix market we reach an average of 90% compression. This is principally due to the fact that the encoding of numbers don't use all the space available (like not using all the bits of a long float) . And also there is a lot of redundant data.

#### 2) Multicast with data compression

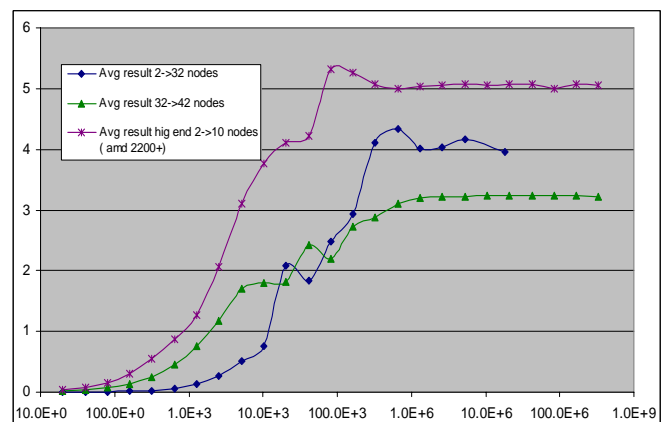

Figure 10. Average Data throughput (Mb/s) Vs matrix size (byte) with 2 nodes up to 42

The graph above (figure 10) shows the throughput of multicast version of MPICH-V2 with data compression. We notice that the data compression gives a tremendous throughput increase compared to the multicast alone (the best is an increase of 720%). This is because the data to send is tremendously reduced and the creation of the FEC data is greatly accelerated due to the simplification of the data by compression and the lack of redundant information.

The performance decrease still exists, but it occurs when we start to transfer larger matrices. In addition the overall decrease is reduced to less than 10%, due to the overall throughput improvement (compared to the 50 % drop seen in the average throughput in figure 7).

We can also see the lack of influence made by the number of receiving nodes in the communication. The only influence comes from the hardware used to perform the tests. We noticed better performance for the first part of the graph (figure 10). That is because we used the AMD 2200+ as source, which made the process of FEC calculation and compression for small matrices faster. In contrast for larger matrices we see that the performance is lower than with the AMD 1800+ as source. Because we used P3 500 as receivers in the blue/circle and green/triangle tests (figure 10), calculation takes more time especially for large file (less memory and CPU power). This leads to an inferior overall performance (- 20%).

We decided when we saw this result to test with only the high-end computers; this experiment gives us the purple/star curve (figure 10). In addition, the overall performance increased by ~25% compared to the 2-32 nodes experiment, blue/circle (figure 10). This means that we can increase overall performance by using a wise choice of source nodes.

We also need to consider the bandwidth usage. With data compression we reduce the amount of data to send to 10% of the size of the matrix (for large matrices). We then need to increase the amount of data to send by 100% to 200%, of the compressed data, for the FEC. These two steps decrease the overall data to send by 70-80% of the initial data. The data resulting from the above two steps is sent on all the channels (in these experiments 5 channels were used). This saves a lot of bandwidth and compensates for the network pollution from the multi layered communication.

We didn't compare the result with compressed P4 or V2 because we didn't want to compare the bandwidth gain in point to point communication with MPI but in 1 to N communications and how we can improve multicast by using this technique in grids. When we compare the throughput we obtain with P4 or V2, we see that we are no longer influenced by the number of nodes. With data compression, we can achieve better throughput than P4 and V2 without compression (but with data compression, P4 and V2 can achieve greater performance easily on a small number of nodes). If we have slower connection this advantage is reduced.

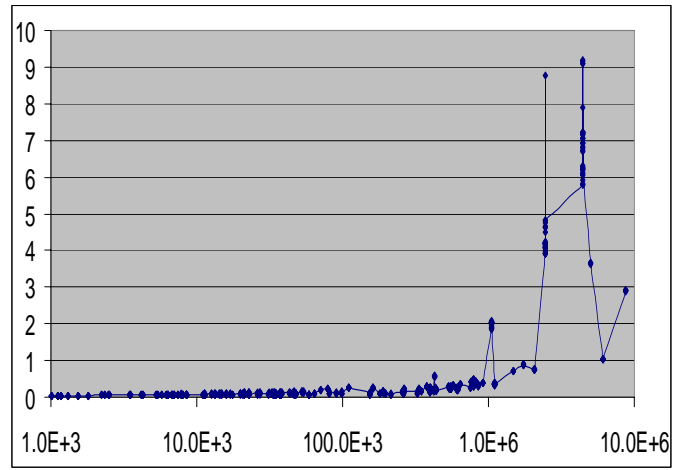

Figure 11. Average time to transfer data (s) Vs matrix size (byte)

We can also see in figure 11 some spikes. They represent similar size of matrices with different data sets (same size but different information). This is because these data sets have different compression ratios. Moreover, the difference in compression ratio directly affects the transfer rate. A variation of compression ratio of 10 % can add more than one second of transfer time.

*3) Fault tolerance*

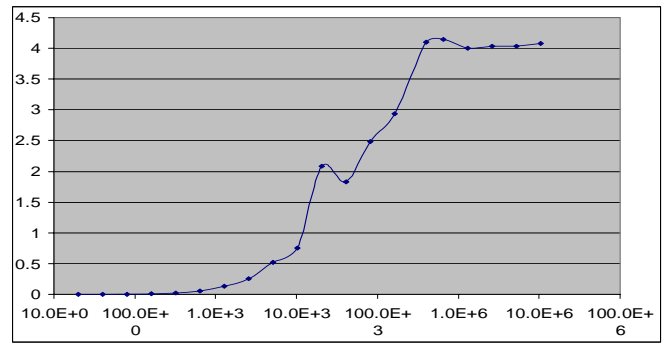

Figure 12. Average Data throughput (Mb/s) Vs matrix size (byte) with 32 nodes with an average packet loss of 7%

We have also simulated packet loss with MCL; we just drop at random a certain amount of packet to send. From 1% to 10%, the impact of this loss is also non-existent thanks to the efficiency of FEC algorithm. Figure 12 shows a throughput with a simulated 7% packet loss on the network. We can notice that the loss of 5% of UDP packets (multicast uses UDP) can represent at the end up to 25% of overall data loss (fragmented packet etc …). We use a loss between 1 and 10%, because it represents the average dropping ratio that exists over internet links without sever perturbation.

*4) Benefits for Grid systems*

Scalability is major benefit of multicast. For the sender it doesn't make any difference if the audience is just three people or three thousand. Indeed, the sender needs to send out *only one copy*. This means: no duplicate packets and a highly efficient bandwidth usage (in theory). The only trade off is latency, when you want to send large chunk of data the latency problem is irrelevant compared to the duration of the transfer. It's easy to change the code to account for the case when the data chunk is too small, and for which unicast transfer should be used.

This responds to a current trend in the Grid community: the need to scale up. To say that such applications stress the capabilities of wide-area networks is an understatement. When we start to have thousand of receiving nodes, unicast is completely useless because of its redundant use of links. To fully utilize the network resources, receivers must be arranged in efficient communication trees. This in turn requires the efficient coordination of a large number of individual components, leading to a concomitant need for resilience to node and link failures. This can be achieved using IP multicast.

ALC is a massively scalable and reliable multicast protocol proposed by the Reliable Multicast Transport (RMT) working group [13] which relies on multiple multicast groups. Using multiple multicast groups enables each receiver to receive data at an appropriate rate by subscribing to as many groups as made possible by their networking (or processing power) capabilities. It provides adaptability for each peer through this multi layered connection. Even if each node has the same bandwidth capabilities, they don't use it the same way (some nodes will have to communicate more and by doing so they have less average bandwidth available for each communication). The use of ALC + FEC provides reliability, which is lacking by design in IP multicast. With an implementation of this protocol we can reach transmissions rates up to 13.6 Mbps between peers over a 100Mbps link.

IV. HARNESSING THE BENEFIT FROM MULTICAST IN A GRID

We have seen that the use of multicast in a Grid through mpich-V2 could be useful especially since the number of receivers does not affect the data throughput. In addition, the use of data compression with Zlib greatly increases the performance of the multicast protocol and reduces the impact of some limitations built in the libraries we used (MCL). By using this combination, we can reach better performance and more efficient use of bandwidth than P4 or V2. However there is a major drawback, multicast needs a stable structure to be efficient. If nodes using multicast come and go, the system will spend its time updating the multicast tree and trying to repair broken communication links. Multicast capability of the routers must be turned on (most of the time ISPs do not do this, and this feature may become a key component in Grid environment), and intelligent switches to avoid network are not cheap. We think multicast can be useful in grids even with the problems inherent in the multicast protocol.

*A. Reality of IP multicast*

There is actually no true reliable multicast protocol (IETF is building one [14]) and congestion control is almost nonexistent. Nevertheless, the introduction of this protocol introduces problems of its own and reduces the scalability (NACK implosion for example). Its use needs support of higher-level functionality from end systems. Routers maintain per-group state and spanning tree associated with each source.

Now the implementations of multicast in the real world are more on the order of One to Many schema and tend to forsake the Many to Many schema. These kinds of messages have a huge impact on the working of the routers and the protocols used to maintain routes and spanning tree for the multicast. In addition, routers have physical characteristics (memory, cpu, etc …) and resulting power and hence price. The cost of newer and more advanced hardware effects their adoption.

Multicast-Tree has several "Single Points of Failure". In the event of a network failure, Reliable Multicast Approaches, (RMTP or Distributed Caching) are limited (e.g. by cache size). Information of changes in the architecture of the network from router to router takes time to propagate. This can be more than 3 minutes for only a few routers. This latency can induce a lot of problems (peer thinking that other peers are down, etc …). In addition, this can impact the integration of clusters using these protocols within a larger environment (GRID).

*B. Architectures that use Multicast*

As we have just seen IP multicast is not really the ideal protocol for Grids, but it can provide a lot of amelioration to the actual way of communicating. For example, a solution to the single point of failure and propagation of information within the network architecture: transporting the same data, on redundant paths simultaneously. A solution can be provided by using scalable overlay architecture as implemented in P2P networks. Nodes in the overlay network act as "Super Nodes" and build up alternate paths for transportation. Using multicast techniques to broadcast data at the higher level is not really feasible at this time. The One to Many mode of communication seems more suited to the transfer of data from super node to normal node within a cluster belonging to the Grid application.

## V. Conclusion and Future Work

In this paper we presented a new implementation of a reliable multicast version of MPI. The performance of this implementation proved that Multicast can be a viable option for large groups of communicating nodes (i.e. Grid). This is despite their volatility and heterogeneous characteristic. It can be used to improve communication especially with broadcast and other form of N to M computers. In conjunction with ACL, which is a good system to handle failure of peers or messages, we can maximize the use of available bandwidth.

In addition, one thing we have to realize is that in Grid we tend to reduce the bandwidth usage to its minimum, because of the lack of it. This is a precious resource, but if it is such a precious resource why don't we use it to its maximum by making program / libraries (etc...) that use the network almost all the time to avoid the loss of such precious bandwidth. We think that a good global management of data to process can achieve better performance by building an intelligent interface that can manage data flows through the Grid. These large data flows have some similarities with large scale P2P. Hence they have similar system problems, especially in the case of linear algebra [15] and multicast can help solve part of this problem.

However, multicast needs a minimum of stability .We know that we can't really provide such stability at the micro level of the architecture (volatility, heterogeneity of the peers), and we cannot provide it at a macro level because of its complexity. However, we can provide it at the medium level. One way to achieve this and it's the subject of future work, is to find a way to construct self-organized cluster of neighbors in a way to optimize the communication but also available resources. Each peer has its own neighbors from which it can optimize message routing, search, resources allocation, etc. This neighborhood will be built depending on the purpose of the network and the goals. By doing so we can use advantages of two different systems: Grid and P2P.

## VI. Acknowledgements

This work was realized as part of the Project: Grand Large [18]. This paper has been developed as a result of collaboration between many individuals and teams. The authors wish to acknowledge contributions from many people, including Franck CAPPELLO, the Xtremweb / MPICH-V team at LRI and MCL team at INRIA. And especially to Liam McDermott who spend a few hours correcting my English for this paper.

### References:

[1] MPICH-V2: a Fault Tolerant MPI for Volatile Nodes based on Pessimistic Sender Based Message Logging Aurelien Bouteiller, Franck Cappello, Thomas Herault, Geraud Krawezik, Pierre Lemarinier, Frederic Magniette *LRI, Universite de Paris Sud, Orsay, France*
[2] http://www.lri.fr/~fedak/XtremWeb/introduction.php3
[3] http://www-unix.mcs.anl.gov/mpi/mpich/
[4] http://www.inrialpes.fr/planete/people/roca/mcl/mcl.html
[5] http://www.ietf.org/rfc/rfc3451.txt
[6] http://www.ietf.org/internet-drafts/draft-ietf-rmt-bb-norm-06.txt
[7] http://www.inrialpes.fr/planete/people/roca/mcl/doc/rfc3453.txt
[8] http://www.cs.wisc.edu/condor/
[9] Reliable multicast fec: where to use FEC (1996) Jörg Nonnenmacher, Ernst W. Biersack Protocols for High-Speed Networks
[10] Evaluating the Utility of FEC with Reliable Multicast , Dan Li and David R. Cheriton , Computer Science, Stanford University
[11] *V. Roca, Z. Khallouf, J. Laboure*, ``Design and Evaluation of a Low Density Generator Matrix (LDGM) large block FEC codec'', Fifth International Workshop on Networked Group Communication (NGC'03), Munich, Germany, September 2003
[12] http://www.gzip.org/zlib/zlib_docs.html
[13] RMT working group http://www.ietf.org/html.charters/rmt-charter.html
[14]http://www.ietf.org/html.charters/rmt-harter.html
[15] Serge Petiton, Lamine Aouad."Large Scale Peer to Peer Performance Evaluations, with Gauss-Jordan Method as an example", in: HeteroPar03 conference, 2003, Septembre, Springer Verlag, LNCSPoland.
[16] http://www-unix.mcs.anl.gov/mpi/mpich/
[17] http://math.nist.gov/MatrixMarket/
[18] Grand Large: http://www.inria.fr/recherche/equipes/grand-large.fr.html